\newcommand{\be}{\begin{equation}}
\newcommand{\ee}{\end{equation}}
\newcommand{\ba}{\begin{array}{c}}
\newcommand{\ea}{\end{array}}
\newcommand{\bqa}{\begin{eqnarray}}
\newcommand{\eqa}{\end{eqnarray}}
\documentclass[
   ,final
  ]
  {aipproc}

\layoutstyle{6x9}
\SetInternalRegister\hbadness{8000} 

\newcommand\doingARLO[2][]{%
  \ifx\mmref\undefined #1\else #2\fi
}

\begin{document}

\title
      {The properties of the $\sigma$ and $\kappa$ resonances in a new unitarization approach}


\author{H.~Q.~Zheng}{
  address={Department of Physics, Peking University, Beijing 100871, China},
  email={zhenghq@pku.edu.cn},
}

\iftrue
\author{Z.~Y.~Zhou}{
  address={Department of Physics, Peking University, Beijing 100871, China},
}

\author{G.~Y.~Qin}{
  address={Department of Physics, Peking University, Beijing 100871, China},
  altaddress={Present address: Department of Physics, McGill University,  Montreal, Canada}
}

\author{Z.~G.~Xiao}{
  address={Department of Physics, Peking University, Beijing 100871, China},
}

\fi

\copyrightyear  {2001}

\begin{abstract}
A new unitarization approach is discussed and applied to study the
elastic $\pi K$ scattering process.
 The existence of the light $\kappa$
resonance is firmly established if the scattering length in the
I=1/2 channel does not deviate too much from its value obtained
from chiral perturbation theory, and a precise determination of
the mass and width of the $\kappa$ resonance requires a precise
determination of the scattering length parameter.
\end{abstract}

\date{\today}

\maketitle

In history the $\sigma$ particle was firstly purposed by
 Gell-Mann and Levy in association with the linear $\sigma$
 model.
Non-linear realization of chiral symmetry was  later
discovered~\cite{Callan} and since then there existed argument
that the $\sigma$ meson is unnecessary for chiral symmetry and
even in contradiction to experiments. However there are also
people, many of them are among the audience, insist on the
existence of $\sigma$ resonance which results in the return of
$\sigma$ in the Review of Particle Properties (named as
$f_0(600)$) after disappearing for about 30
years.~\cite{Tornqvist} The postulated $\kappa$ resonance also has
a rather long history~\cite{Jaffe1977} and the status is even more
intriguing despite the recent experimental results from the E791
and the BES Collaborations.~\cite{E791BESkappa} The dynamics
related to the $\sigma$ and $\kappa$ resonance is of highly
non-perturbative, strong interaction nature. Since it is always
not easy to separate a distant pole from the background
contributions, conclusions found in the literature are very often
model dependent.

However, in Ref.~\cite{XZ00,HXZ}, a model independent dispersive
analysis has been proposed and it is demonstrated that the
non-linear realization of chiral symmetry, or chiral perturbation
theory ($\chi$PT) actually needs the $\sigma$ resonance to
accommodate for the experimental data. The crucial point to attack
the problem is to consider the dispersion relations for the
following two quantities,
\begin{eqnarray}
& & F(s)=\frac{1}{2i\rho(s)}\big(S(s)-\frac{1}{S(s)}\big)\ ,\,\,\,
\tilde{F}(s)=\frac{1}{2}\big(S(s)+\frac{1}{S(s)}\big)\ ,
\end{eqnarray}
where $\rho(s)$
is the kinematic factor.
 The functions $\rho(s)F(s)$ and
$\tilde{F}(s)$ are respectively the analytic continuation of the
imaginary and real part of the partial wave $S$-matrix defined in
the elastic region and they satisfy the following dispersion
relations:
 \bqa
  F(s)&=& \alpha-\sum_j{1/ (2i\rho(z^{II}_j))\over
 S'(z^{II}_j)(s-z^{II}_j)}  +{1\over\pi}\int_L{{\rm Im}_LF(s')
  \over s'-s} ds'+{1\over\pi}\int_R{{\rm Im}_RF(s')
  \over s'-s} ds',\nonumber \\
 \tilde{F}(s)&=&\tilde{\alpha}+\sum_j{1\over
 2S'(z^{II}_j)(s-z^{II}_j)}
+{1\over\pi}\int_L {{\rm
 Im}_L\tilde F(s')
 \over s'-s} ds' 
 +{1\over\pi}\int_R {{\rm
 Im}_R\tilde F(s')
 \over s'-s} ds'\ . \label{cos2d}
\eqa
   In the above expressions the subscript $L$ represents dynamical
   cuts rather than the physical right hand cuts. $R$ represents right
   hand cuts starting from the second physical threshold. The left hand cut
   is in general rather complicated but for equal mass scatterings
   like $\pi\pi$ scatterings the situation is much simplified:
    $L=(-\infty,0]$ and $R=[4m_K^2,\infty)$. One subtraction to the
    dispersion integrals is understood. In Eq.~(\ref{cos2d})
    $\alpha$ and $\tilde\alpha$ are  subtraction constants and
   $z_j$ denotes  pole positions on the second sheet. When $z_j$ is real it represents a virtual state
   pole, when $z_j$ is complex it must appear in one conjugate pair together with $z_j^*$, representing a resonance.
The experimental curve of the function $F$ is convex, yet chiral
perturbation theory predicts
   a negative and concave left hand integral contribution.~\cite{XZ00} This fact unambiguously
establishes the existence of the $\sigma$ resonance, if the chiral
prediction to the cut integral is qualitatively correct. For more
details we refer to Ref.~\cite{XZ00}.

\begin{figure}%
\includegraphics[height=.25\textheight]{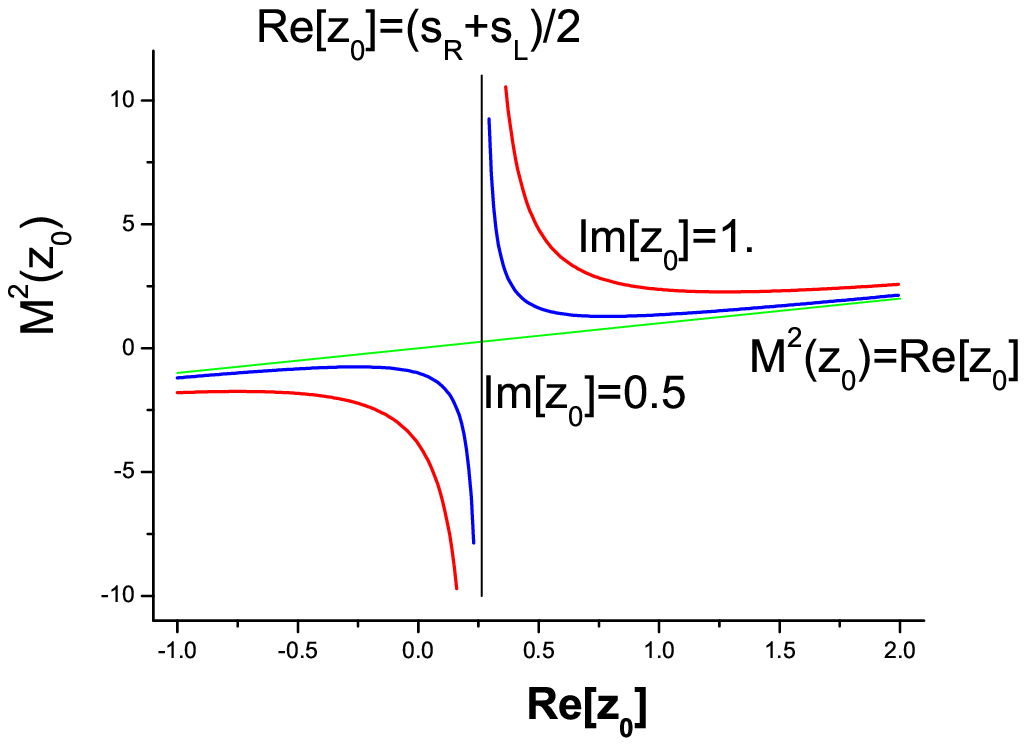}
\includegraphics[height=.25\textheight]{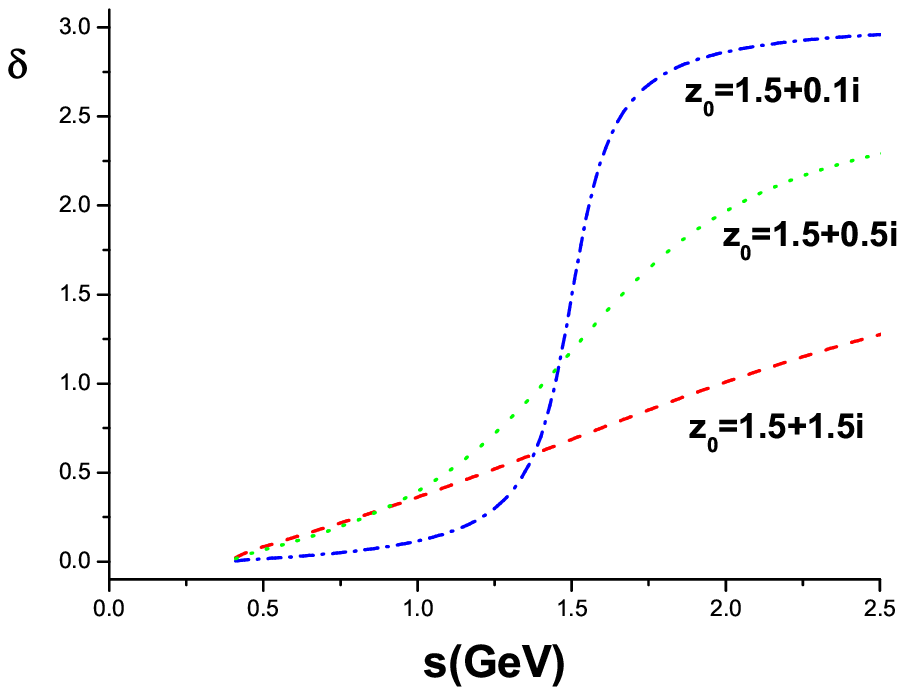}
\caption{\label{r0 function} The left figure shows $M^2(z_0)$ as a
function of $\mathrm{Re}[z_0]$, for fixed $\mathrm{Im}[z_0]$. The
right figure give some examples of resonances and their
contributions to the phase shift.}
\end{figure}

From Eq.~(\ref{cos2d}) one obtains the generalized unitarity
condition which holds on the entire complex $s$ plane~\cite{HXZ}:
 \be
  \tilde F^2+ (\rho F)^2=1\ .
 \ee
This equation is used to obtain solutions of the simplest $S$
matrices. Here `simplest' means those solutions of unitary $S$
matrices contain no cut integrals as appeared in Eq.~(\ref{cos2d})
and contain minimal set of poles, i.e., one or two. The one pole
solution represents a virtual/bound state whereas the two pole (on
the second sheet) solution represents a resonance. The solution
representing a resonance located at $z_0$ (having positive
imaginary part) and $z_0^*$ for un-equal mass scatterings is the
following:
\bqa\label{a pair resonaces S matrix}%
S(s) = \frac{{M^2}(z_0)-s + i\rho(s)s G[z_0] }
  {{M^2}(z_0)-s - i\rho(s)s G[z_0]}\ ,
\eqa where
 \bqa {M^2}(z_0) = \mathrm{Re}[z_0] +
\frac{\mathrm{Im}[z_0]\,
     \mathrm{Im}[z_0\,\rho (z_0)]}{\mathrm{Re}[
     z_0\,\rho (z_0)]}\ ,\,\,
G[z_0] =\frac{\mathrm{Im}[z_0]}{\mathrm{Re}[z_0\,\rho (z_0)]}\ .
\eqa
The Eq.~(\ref{a pair resonaces S matrix}) is very interesting as
it reveals the remarkable difference between a narrow resonance
located far above the threshold and a light and broad resonance.
In fact, $s=M^2(z_0)$ is the place where the resonance
contribution to the phase shift passes $\pi/2$. However, a light
and broad resonance corresponds to a very large $M(z_0)^2$. When
$\mathrm{Re}[z_0]\le (s_L+s_R)/2$ ($s_L=(m_K-m_\pi)^2$,
$s_R=(m_K+m_\pi)^2$), the phase shift never reaches $\pi/2$! See
fig.~\ref{r0 function} for more illustrations.

It is worthwhile to make a pedagogical analysis to a widely used
parameterization form found in the literature:
 {\be\label{mBW}
S=\frac{M^2-s+i\rho(s)g}{M^2-s-i\rho(s)g}\ .\ee}
 For a
sufficiently large $M^2$ and small $g$ and for equal mass
scatterings,  such an $S$ matrix contains a resonance and a
\textit{virtual} state~\cite{zhengtalk03}. However, the latter is
not predicted by $\chi$PT and violates the validity of chiral
expansion at the pole position and therefore should be abandoned.
For unequal mass scattering the $S$ matrix Eq.~(\ref{mBW})
contains two resonance poles. The situation is depicted in
fig.~\ref{two pairs of resonances}. The virtual state pole in
$\pi\pi$ scattering and the resonance pole in $\pi K$ scattering
have a common origin: they are both generated from $s=0$ due to
the kinematical singularity of $\rho(s)$ at $s=0$ and should
therefore be removed on the same footing. Moreover, the physical
pole contribution and the spurious pole contribution to the
scattering length are additive and are both positive. The
accompanied spurious pole can have a larger contribution than the
resonance contribution itself if the resonance is light and broad!
It is therefore incorrect to use Eq.~({\ref{mBW}) to discuss a
light and broad resonance.
\begin{figure}%
\includegraphics[height=.25\textheight]{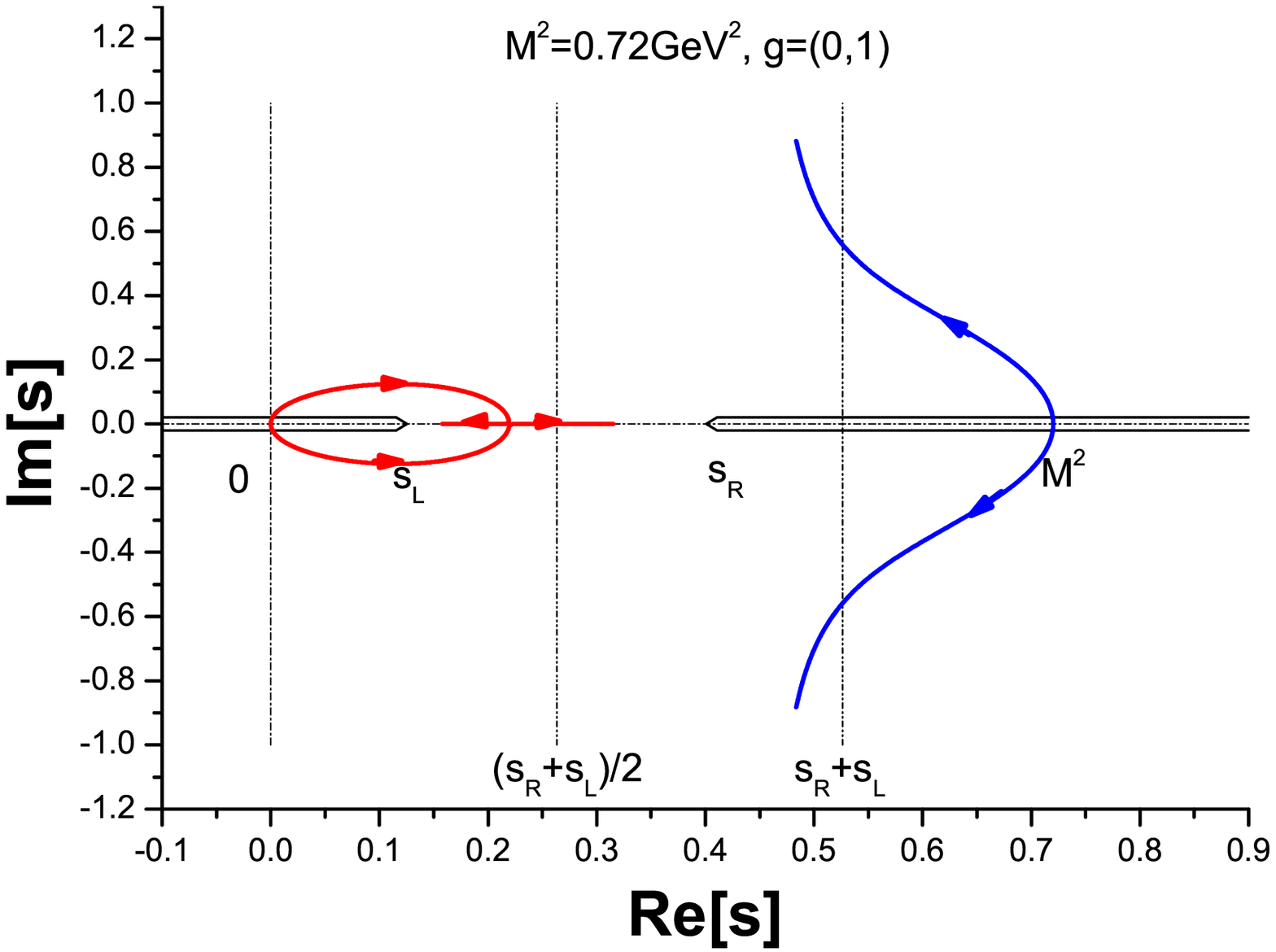}
\includegraphics[height=.25\textheight]{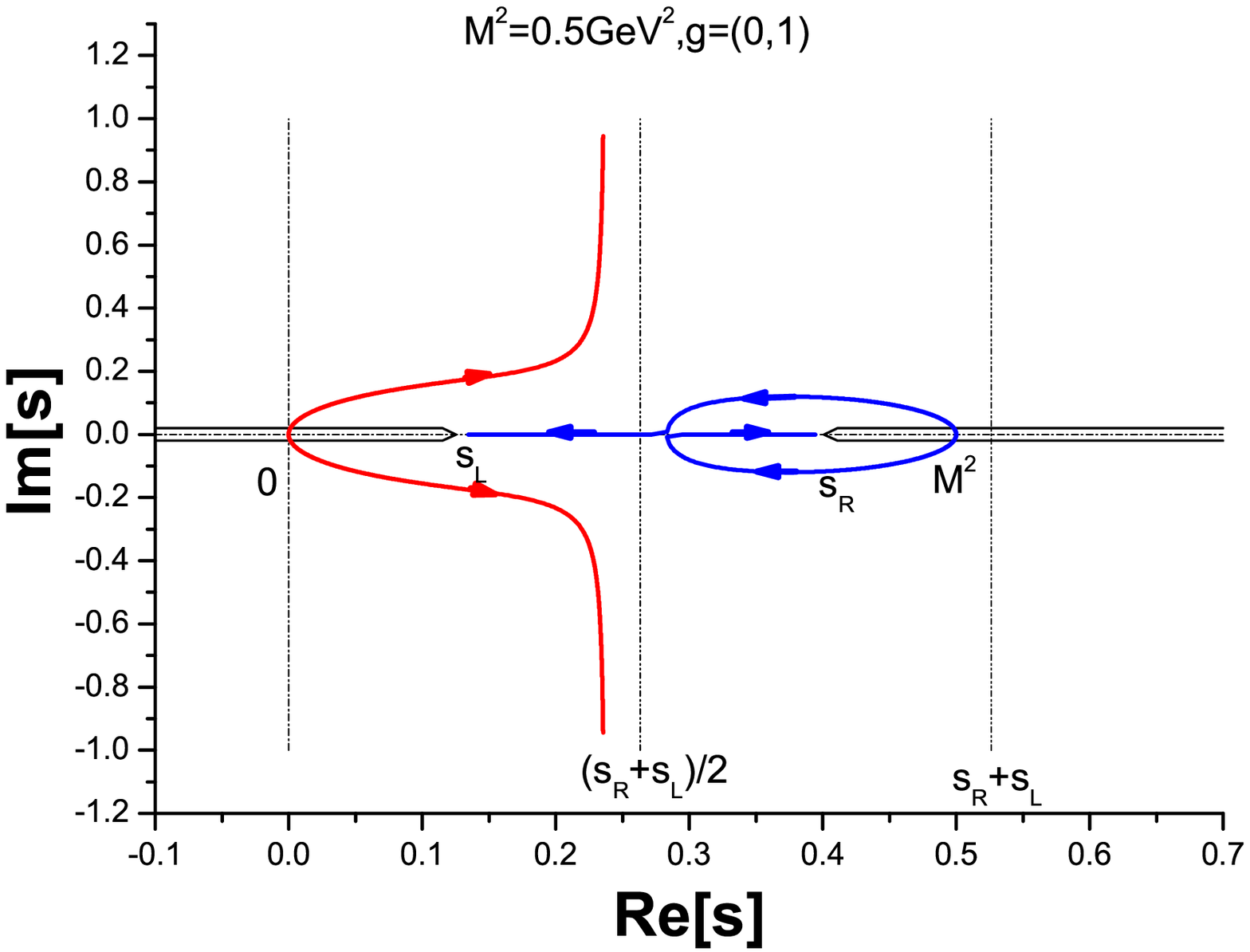}
\caption{\label{two pairs of resonances} The traces of two pairs
of resonances from Eq.(\ref{mBW}) when increasing $g$ for
different $M^2$. We give two typical figures: left)
$M^2>(s_R+s_L)$; right) $M^2<(s_R+s_L)$.}
\end{figure}%

Since a unitary matrix divided by any unitary matrix is still
unitary. If we single out all  poles of an $S$ matrix, we have
without any loss of generality the following form:{ \be
 S^{phy.}=\prod S^{poles}\cdot S^{cut}\ .
\ee} where $S^{cut}$ no longer contains any pole and it can be
parameterized as:
 { \bqa
 S^{cut}&=&e^{2i\rho f(s)}\ ,\nonumber\\
 f(s)&=&f_0+\frac{s-s_0}{\pi}\int_{L}\frac{{\rm
 Im}_Lf(s')}{(s'-s_0)(s'-s)}
 +\frac{s-s_0}{\pi}\int_{R}\frac{{\rm Im}_Rf(s')}{(s'-s_0)(s'-s)}\
 .
\eqa
It can be demonstrated that the discontinuity of $f$ obeys the
following simple relations:~\cite{tocome} { \bqa\label{discf}
  \mathrm{disc}\,f&=&\mathrm{disc}\{\frac{1}{2i\rho(s)}\log
  \left[S^{phy}(s)\right]\}\ .\eqa
The equation (\ref{discf}) is  useful when  estimating the
background contributions. For example we can simply  approximate
$S^{phy}$ in Eq.~(\ref{discf}) by $S^{\chi \mathrm{PT}}$ on $L$,
 {\be\label{discf'}
  \mathrm{disc}f_{ L}=\mathrm{disc}\{\frac{1}{2i\rho(s)}\log
  \left[S^{\chi\mathrm{PT}}(s)\right]\}\ ,
  \ee}
to estimate the background contributions.

 Now it is at the stage
to use the newly established unitarization scheme to study the
$\pi K$ scatterings. In here $R$ in principle starts from
$(m_K+m_\eta)^2$ to $+\infty$ but this cut is rather weak in
practice and therefore it is appropriate to take
$R=[(m_K+m_\eta')^2,+\infty)$. We  fit the LASS data~\cite{LASS}
up to 1430MeV (about 20 MeV below the $K\eta'$ threshold). There
are totally six parameters $a_0^{1/2}$, $a_0^{3/2}$, two pole
parameters for $\kappa$ and two for $K^*(1430)$. It is found
appropriate to truncate the left hand integral at
$\Lambda^2_{L,I=1/2}\simeq\Lambda^2_{L,I=3/2}\simeq 1.5$GeV$^2$
(since the once subtracted dispersion integral is convergent the
numerical output is actually not very sensitive to the truncation)
and we obtain:
 { \bqa\label{resI}
&&\chi^2_{d.o.f.}=38.35/(60-6)\ ;\nonumber\\
&&M_\kappa=594\pm 79MeV\ ,\,\,\, \Gamma_\kappa=724\pm 332MeV\ ;\nonumber\\
&&a_0^{1/2}=0.284\pm 0.089\ ,\,\,\,a_0^{3/2}=-0.129\pm 0.006\ ;\nonumber\\
&&M_{K^*}=1456\pm 8MeV\ ,\,\,\, \Gamma_{K^*}=217\pm 31MeV\ .
 \eqa}
 The fit quality of the above results are rather good though
 the width of the $\kappa$ resonance is quite flexible.
 However the fit prefers a somewhat larger scattering length in the I=1/2
  channel and much larger (in magnitude) scattering length
  parameter in the I=3/2 channel comparing with the
  $\chi$PT predictions to the $\pi K$ scattering lengths~\cite{Meissner et al.}:
  $a_0^{1/2}=0.18\pm 0.02$, $a_0^{3/2}=-0.05\pm 0.02$.
It is necessary to further clarify the issue whether the $\kappa$
resonance exist. According to Pennington, a resonance exist if
without it the total $\chi^2$ of the fit increase significantly.
If we freeze the $\kappa$ resonance, we get,
$\chi^2_{d.o.f.}=63.67/(60-4)$, $a_0^{1/2}=0.446\pm 0.006$ and
other outputs are similar to Eq.~(\ref{resI}). Comparing with the
results in Eq.~(\ref{resI}) the $\chi^2_{d.o.f.}$ given in this
way is increased by a factor of 1.7. If this is still not enough
to support the existence of $\kappa$, the value of $a_0^{1/2}$
given by freezing the $\kappa$ degree of freedom is too large
comparing with the $\chi$PT value: the fit value of $a_0^{3/2}$ is
about  4 $\sigma$ away from $\chi$PT result whereas $a_0^{1/2}$ is
14 $\sigma$ away! The conclusion is that if $a_0^{1/2}$ does not
deviate much from its value predicted by $\chi$PT then $\kappa$
resonance must exist. On the contrary, it is estimated that if
$a_0^{1/2}$ becomes larger than roughly 0.35 then the existence of
the $\kappa$ resonance becomes doubtful.

It is sometimes found in the literature the discussions on the
constraint of Adler zero on the scattering amplitude. In our
scheme, it is possible to embed this constraint into our
parameterization form. In fact, the Adler zero automatically
emerges in our approach if the subtraction constant $f_0$ is
limited within certain range. This is because all the
resonance(and virtual state) $S$ matrices are real and less than 1
when $s_L<s<s_R$. On the contrary the cut integrals contribute a
factor larger than 1, therefore the $T$ matrix zero is obtainable
in the right place when $f_0$ is confined to a certain range,
which in turn put some constraints on the magnitude of the
scattering length parameter itself. For example, within the range
$0<a_0^{1/2}<0.26$ there exists a $T$ matrix zero in the region
$s_L<s<s_R$, and when $a_0^{1/2}\simeq 0.20$ the zero locates in
the place close to the one loop $\chi$PT prediction. We make
further fit by confining $a_0^{1/2}$ in the region $0.18\pm 0.02$
(other conditions are the same as the fit described in above) and
the results follow: \bqa\label{resI}
&&\chi^2_{d.o.f.}=38.96/(60-6)\ ;\nonumber\\
&&M_\kappa=646\pm 7MeV\ ,\,\,\, \Gamma_\kappa=540\pm 42MeV\ ;\nonumber\\
&&a_0^{1/2}=0.2\ ,\,\,\,a_0^{3/2}=-0.128\pm 0.006\ ;\nonumber\\
&&M_{K^*}=1450\pm 5MeV\ ,\,\,\, \Gamma_{K^*}=232\pm 25MeV\ .
 \eqa
The Adler zero position is now at $s_A\simeq 0.245$GeV. If in this
fit we further freeze the $\kappa$ degrees of freedom we would
obtain a total $\chi^2$ $\sim 750$. This clearly demonstrates the
existence of the $\kappa$ resonance. For more details of our
analysis we refer to Ref.~\cite{tocome}.


\begin{thebibliography}{99}

\bibitem{Callan}
S.~R.~Coleman, J.~Wess and B.~Zumino, Phys. Rev. {\bf 177}
(1969)2239; C.~G.~Callan, S.~R.~Coleman, J.~Wess and B.~Zumino,
Phys. Rev. {\bf 177}(1969)2247.
\bibitem{Tornqvist}See for example, N.~A.~Tornqvist,
Invited talk at YITP Workshop on Possible Existence of the sigma
meson and its Implications to Hadron Physics, Kyoto, Japan, 12-14
Jun 2000, hep-ph/0008136.
\bibitem{Jaffe1977}
R.~Jaffe, Phys. Rev. {\bf D15}(1977)267. \\
 Recent discussions may be
 found in, S.~Ishida $et$ $al.$, Prog. Theor. Phys.{\bf
98}(1997)621;
\\
S.~N.~Cherry and M.~R.~Pennington, Nucl. Phys. A688(2001)823.
\bibitem{E791BESkappa}E.~M.~Aitala $et$ $al.$(E791 Collaboration), Phys. Rev. Lett.{\bf 89}(2002)121801;\\
J.~Z.~Bai $et$ $al.$(BES Collaboration), hep-ex/0304001.
\bibitem{XZ00}
Z.~G.~Xiao and H.~Q.~Zheng, Nucl. Phys. {\bf A695}(2001)273.
\bibitem{HXZ}
J.~Y.~He, Z.~G.~Xiao and H.~Q.~Zheng, Phys. Lett. {\bf
B526}(2002)59; Erratum: $ibid$. {\bf B549}(2002)362.
\bibitem{zhengtalk03}H.~Q.~Zheng,
Talk given at International Symposium on Hadron Spectroscopy,
Chiral Symmetry and Relativistic Description of Bound Systems,
Tokyo, Japan, 24-26 Feb 2003, hep-ph/0304173.
\bibitem{tocome}
H.~Q.~Zheng $et$ $al.$, to appear.
\bibitem{LASS}D.~Aston $et$ $al.$ (LASS Collaboration), Nucl. Phys. {\bf B296}(1988)493.
\bibitem{Meissner et al.}
Ulf-G.Meissner, Nucl. Phys. {\bf B357}(1991)129.
\end{thebibliography}

\end{document}